\documentclass[10pt,journal]{IEEEtran}

\ifCLASSINFOpdf
\else
\fi

\usepackage{amsmath}
\usepackage{graphicx}
\usepackage{diagbox}
\usepackage{multirow}
\usepackage{threeparttable}
\usepackage{cite}
\usepackage{amsfonts,amssymb}
\usepackage{color}
\usepackage{amsmath}
\usepackage{graphicx} 
\usepackage{epstopdf}

\usepackage{xcolor}
\usepackage{xpatch}
\makeatletter
\def\changeBibColor#1{%
	\in@{#1}{9113273,Moose328961,10.5555/3294673}
	\ifin@\color{black}\else\normalcolor\fi	
}
\xpatchcmd\@bibitem
{\item}
{\changeBibColor{#1}\item}
{}{\fail}
\xpatchcmd\@lbibitem
{\item}
{\changeBibColor{#2}\item}
{}{\fail}
\makeatother

\begin{document}
%
\title{High-Performance Low-Complexity Hierarchical Frequency Synchronization for Distributed Massive MIMO-OFDMA  Systems}
%
%
%

\author{Xiao-Yang~Wang, Shaoshi~Yang, Tian-Hao~Yuan, Hou-Yu Zhai, Jianhua~Zhang, and Lajos~Hanzo
\thanks{
		Copyright (c) 2015 IEEE. Personal use of this material is permitted. However, permission to use this material for any other purposes must be obtained from the IEEE by sending a request to pubs-permissions@ieee.org.
	
		This work was supported in part by Beijing Municipal Natural	Science Foundation under Grant L202012 and Grant Z220004, by the Fundamental Research Funds for the Central	Universities under Grant 2020RC05, by National Science Fund of China for Distinguished Young Scholars (Grant No. 61925102), by the Engineering and Physical Sciences Research Council projects EP/W016605/1 and EP/X01228X/1, and by European Research Council's Advanced Fellow Grant QuantCom (Grant No. 789028). (\textit{Corresponding author: Shaoshi Yang}.)
	
X.-Y. Wang, S. Yang, T.-H. Yuan, H.-Y. Zhai are with School of Information and Communication Engineering, Beijing University of Posts and Telecommunications, and the Key Laboratory of Universal Wireless Communications, Ministry of Education, 
Beijing 100876, China (e-mail: \{wangxy\_028, shaoshi.yang, yth\_97, 2hy\}@bupt.edu.cn).

J. Zhang is with School of Information and Communication Engineering, Beijing University of Posts and Telecommunications, and the State Key Laboratory of Networking and Switching Technology, Beijing 100876, China (e-mail: jhzhang@bupt.edu.cn).

L. Hanzo is with the School of Electronics and Computer Science, University of Southampton, UK (e-mail: lh@ecs.soton.ac.uk). }}

\markboth{IEEE Transactions on Vehicular Technology, submitted, Oct. 2022}%
{Shell \MakeLowercase{\textit{et al.}}: Bare Demo of IEEEtran.cls for IEEE Journals}
%



\maketitle

\begin{abstract}
We propose a high-performance yet low-complexity hierarchical frequency synchronization scheme for orthogonal frequency-division multiple-access (OFDMA) aided distributed massive multi-input multi-output (MIMO) systems, where multiple carrier frequency offsets (CFOs) have to be estimated in the uplink. To solve this multi-CFO estimation problem efficiently, we classify the active antenna units (AAUs) as the master and the slaves. Then, we split the scheme into two stages. During the first stage the distributed slave AAUs are synchronized with the master AAU, while the user equipment (UE) is synchronized with the closest slave AAU during the second stage. The mean square error (MSE) performance of our scheme is better than that of the representative state-of-the-art baseline schemes, while its computational complexity is substantially lower. 
\end{abstract}


%
\IEEEpeerreviewmaketitle

\vspace{-0.5cm}
\section{Introduction}

\IEEEPARstart{M}{assive} multi-input multi-output (MIMO) schemes have widely been viewed as pivotal components of potent mobile communication systems, as a benefit of their remarkable capability of improving the spectral efficiency and energy efficiency~\cite{9113273}. In the traditional centralized massive MIMO, all the antennas are located in a single fixed place, which causes a grave challenge for guaranteeing the transmission rates of user equipment (UE) roaming in the cell-edge areas, due to severe inter-cell interference. Moreover, in the centralized massive MIMO the high-mobility users tend to suffer from frequent handover, which increases both the service outage risk and the signaling cost. As a potential remedy, distributed massive MIMO schemes, such as the cell-free massive MIMO \cite{Ngo2017cell}, has attracted considerable attention thanks to its great potential of reducing the interference and improving the coverage. The classic orthogonal frequency-division multiple-access (OFDMA) technique has become ubiquitous, but it is sensitive to the carrier frequency offsets (CFOs). This issue becomes even more severe in distributed system architectures. The inaccurate compensation of CFOs destroys the orthogonality among subcarriers, thus resulting in severe inter-carrier and inter-user interference~\cite{8676341}. 

The state-of-the-art CFO estimation algorithms for OFDMA systems can be divided into two categories: pilot-aided and blind. The former typically have lower computational complexity and spectral efficiency, since they rely on training sequences \cite{8295107}. By contrast,  the blind schemes in general have higher complexity and spectral efficiency \cite{zhang2016blind}. Both types of algorithms can be applied to single-antenna and  distributed massive MIMO systems. However, the large number of independent oscillators used in distributed massive MIMO systems result in a great many CFOs to be estimated in the uplink. For instance, upon assuming $M$ distributed active antenna units (AAUs) and $K$ UEs, the complexity of CFO estimation is roughly $(M\times K)$ times that of the single-AAU-single-UE system. To reduce the complexity, the authors of \cite{feng2018frequency} adapted the CFO estimation scheme of \cite{zhang2014blind} to the distributed antenna system, where the remote radio units (RRUs) are divided into multiple pairs and the two CFOs in each pair are jointly estimated. Nonetheless, this scheme needs one-dimensional search to estimate the pairwise CFOs and thus imposes excessively high complexity.

In this paper, we extend the concept ``antenna pairing" of \cite{feng2018frequency} into ``AAU grouping", where the AAUs in a single group firstly achieve frequency synchronization with each other by using an algorithm similar to those of  \cite{scalable6760595,Airshare7218555} and then each UE synchronizes with its nearest AAU. The authors of  \cite{scalable6760595,Airshare7218555} only studied synchronization among distributed access points (APs) for supporting distributed coherent transmission in the downlink, but the CFOs between UEs and AAUs were not considered. By contrast, we consider the more challenging ``double-tier" synchronization problem involving asynchronous distributed AAUs and UEs, and propose an efficient hierarchical-architecture based synchronization scheme, where the number of CFOs to be estimated is substantially reduced. Our theoretical analysis and numerical simulations demonstrate that the proposed scheme achieves considerable performance improvement at a lower computational complexity. This is because in our scheme the UEs only have to synchronize with its nearest slave AAU, and all the slave AAUs synchronize with the master AAU. Thus, our scheme enjoys an enhanced signal-to-noise ratio (SNR) in the synchronization process and promises notable mean squared error (MSE) improvement. Furthermore, we demonstrate that the increased mobility of UEs degrades the MSE, but when the mobility is sufficiently low, e.g., less than 10m/s, the influence is marginal. The benefits of our scheme are obtained at the cost of increased challenges in designing the radio access protocols, because additional signaling overhead and radio resource allocation functions have to be introduced in support of the direct communication links between AAUs.  

\section{System Model and Problem Formulation}
\begin{figure}[tbp]
	\centering
	\includegraphics[width=8.2cm,height=3.7cm]{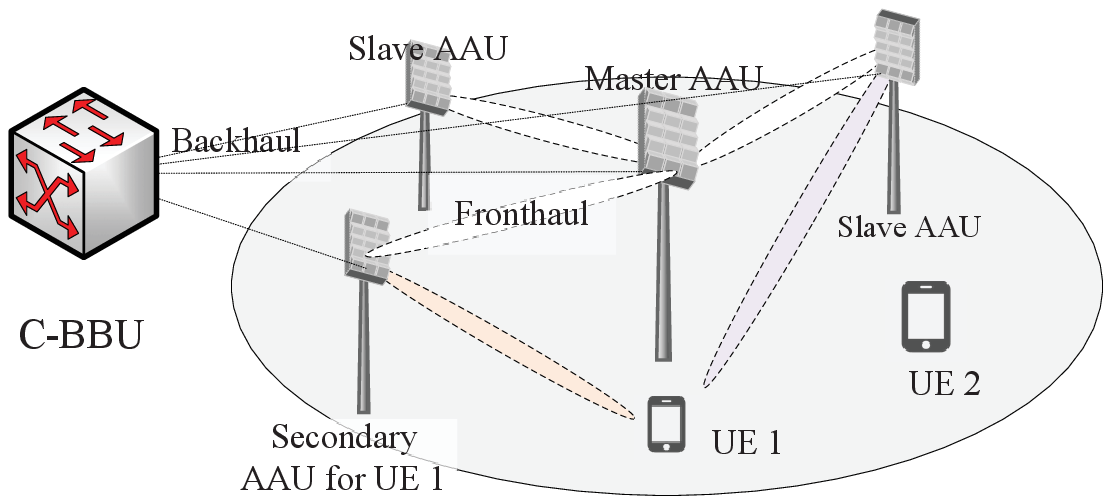}
	\caption{Distributed massive MIMO-OFDMA system model.}\label{system architecture}
\end{figure}
As illustrated in Fig.~\ref{system architecture}, we consider the uplink of a distributed massive MIMO-OFDMA system\footnote{\textcolor{black}{The reference signals that can be used for synchronization, such as those of 5G new radio (NR), are UE-specific and they are not used in a space-division multiple-access (SDMA) style (See Chapter 8 of  \cite{10.5555/3294673}). In fact, the synchronization signals of different UEs are typically used in OFDMA-style, while the data-bearing signals can be transmitted in SDMA-style.}}, which is composed of $K$ single-antenna UEs, $M$ single-antenna AAUs, and a central baseband unit (C-BBU). The AAUs are typically connected to the C-BBU via optic fibre. Each of the AAUs and UEs has a stand-alone local oscillator whose frequency may slightly deviate from the nominal value. Denote the normalized CFO between the $m$th AAU and the $k$th UE by $\varphi^{k}_{m}$. Assume that the estimate of the CFO has to have a reasonable accuracy within $(-0.5,0.5)$, and that the AAUs can communicate with each other either directly or through the C-BBU.

Let us denote the propagation channel from the $k$th UE to the $m$th AAU as ${\boldsymbol{\bf h}^{(k)}_{m} = [h^{(k)}_{m}(1),h^{(k)}_{m}(2),...,h^{(k)}_{m}({L})]^{T}}$, where $L$ represents the number of \textcolor{black}{channel taps}, and the  elements of $\boldsymbol{\mathrm h}^{(k)}_{m}$ are independent and identically distributed (i.i.d) complex Gaussian variables with zero mean and variance $\mathrm{E}[|h^{(k)}_{m}(l)|^{2}] = \frac{1}{ML}$. The total channel gain from the $k$th UE to all AAUs is normalized as $\mathrm{E}[\sum_{m=1}^{M}\sum_{l=1}^{L}|h^{(k)}_{m}(l)|^{2}] = 1$. We also assume that both the channel response and the CFO remain constant during the period of $L_{b}$ OFDM symbols.

Furthermore, the total number of subcarriers is $N$, and the $k$th UE is assigned $N_{k}$ subcarriers. Assume that $P_k$ out of $N_{k}$ subcarriers are occupied by data, while the remaining $N_{k}-P_k$ are reserved as null subcarriers. For the $k$th UE, to characterize the impact of different assignments of the subcarriers bearing data, we denote the mapping matrix between the data and the data-bearing subcarriers as $\boldsymbol{\Gamma}_{k} \in \mathbb{R}^{P_k \times N}$, where the $\left[p,\boldsymbol{\gamma}_{k}(p)\right]$th 
element is set to one, while the other elements are zero, and $\boldsymbol{\gamma}_{k}=[\boldsymbol{\gamma}_{k}(1),...,\boldsymbol{\gamma}_{k}(p),...,\boldsymbol{\gamma}_{k}({P_k})]$ is a vector whose elements are the indices of data-bearing subcarriers of the $k$th UE and these indices are ranked in ascending order, and $p=1,...,P_k$. Moreover, we assume that the length of the cyclic prefix is $N_{c}$ samples.

In an OFDMA system, we denote the $g$th transmitted data vector of the $k$th UE on multiple subcarriers as $\boldsymbol{\rm s}_{g}^{k} = [s_{g}^{k}(1),s_{g}^{k}(2),...,s_{g}^{k}(P_k)]$, and ${\rm E}[s_{g}^{k}(p)] = 1$. The $g$th transmitted data vector after applying the mapping between the data and data-bearing subcarriers can be written as ${ \boldsymbol{\mathrm X}_{g}^{k}=\mathrm{diag}\{\boldsymbol{\rm s}_{g}^{k}\boldsymbol{\Gamma}_{k}\}}$, where $\mathrm{diag}\{\boldsymbol{\rm s}_{g}^{k}\boldsymbol{\Gamma}_{k}\}$ is a diagonal matrix with $\boldsymbol{\rm s}_{g}^{k}\boldsymbol{\Gamma}_{k}$ being its diagonal elements. Let $\boldsymbol{\rm F}$ represent the $N$-dimensional discrete Fourier transform (DFT) matrix and $\eta_{g}(\varphi^{k}_{m})$ be the cumulative phase offset, caused by CFO between the $k$th UE and the $m$th AAU, before the $g$th OFDM symbol emerges. The CFO also causes different phase offset for every sample in the $g$th OFDM symbol. We define the phase rotation matrix between the $m$th AAU and the $k$th UE as follows:
\begin{equation}
\boldsymbol{\mathrm D}(\varphi^{k}_{m}) = \mathrm{diag}(1,e^{j\frac{2\pi \varphi^{k}_{m}}{N}},\cdots,e^{j\frac{2\pi (N-1)\varphi^{k}_{m}}{N}}).
\end{equation}
	
Corresponding to the $g$th OFDM symbol, the \textcolor{black}{time-domain} signal received by the $m$th AAU is represented as\footnote{\textcolor{black}{This model can describe either SDMA or OFDMA systems, depending on the specific value of the UE-to-subcarrier mapping matrix $\boldsymbol{\Gamma}_{k}$.}}
\begin{equation}\label{y_mg}
{\bf y}_{m,g} =\sqrt{N}\sum_{k=1}^{K} \eta_{g}(\varphi^{k}_{m}) \boldsymbol{\mathrm D}(\varphi^{k}_{m}) \boldsymbol{\mathrm F}^{\textrm{H}}\boldsymbol{\mathrm X}_{g}^{k}\boldsymbol{\mathrm F}_{L}\boldsymbol{\mathrm h}_{m}^{k} + \boldsymbol{\mathrm w}_{m,g}, 
\end{equation}
where $ \boldsymbol{\mathrm F}_{L} \in \boldsymbol{\mathbb{C}}^{ N \times L}$ denotes the first $L$ columns of $\boldsymbol{\rm F}$, $\boldsymbol{\mathrm F}^{\textrm H}$ is the conjugate transpose of $\bf F$, and $\boldsymbol{\rm w}_{m,g} \in \boldsymbol{\mathbb{C}}^{ N \times 1}$ denotes the additive white Gaussian noise (AWGN) having the covariance matrix $\mathrm{E}[\boldsymbol{\mathrm w}_{m,g}\boldsymbol{\mathrm w}_{m,g}^{\textrm H}] = \delta_{w}^{2} \boldsymbol{\mathrm I}_{N}$.\footnote{\textcolor{black}{By transforming (2) into the frequency-domain via $\boldsymbol{\mathrm F}{\bf y}_{m,g}$,  we obtain $\boldsymbol{\mathrm F}\boldsymbol{\mathrm D}(\varphi^{k}_{m})\boldsymbol{\mathrm F}^{\textrm{H}}$ as the inter-carrier interference (ICI) matrix \cite{zhang2016blind,feng2018frequency}.}} 

Our goal is to obtain a high-accuracy estimate $\hat{\varphi}^{k}_m$ for any $k$ and $m$ at the lowest possible computational complexity.

\section{The Proposed Hierarchical Frequency Synchronization Scheme}


To demonstrate the feasibility of simplifying the multi-CFO estimation problem relying on a hierarchical architecture, we collectively define $\boldsymbol{\mathrm H}^{k}$ as the channel response between the $k$th UE and $M$ AAUs.\footnote{Note that $\boldsymbol{\mathrm H}^{k}$ does not have to be known at any individual AAU for executing the proposed synchronization algorithm.} Thus we have
\begin{equation}
{{\bf H}^{k} = [{\bf h}_1^{k},{\bf h}_2^{k},\cdots,{\bf h}_M^{k}]^{\textrm T} \in {\mathbb{C}}^{M \times L}}, 
\end{equation}
where $(\cdot)^{\textrm T}$ stands for the transpose of a matrix. 

Then, the $(M \times N)$ received signal matrix of all the $M$ AAUs can be formulated as
\begin{equation}
\begin{aligned}
\boldsymbol{\mathrm Y}_g & =[{\bf y}_{1,g},{\bf y}_{2,g},\cdots,{\bf y}_{M,g}]^{\textrm T} \\ 
& =\sqrt{ N}\sum_{k=1}^{ K} (\eta_{g}(\varphi^{k}_{m})\boldsymbol{\bf H}^{k}\boldsymbol{\bf F}_{L}^{\textrm T}\boldsymbol{\bf X}_{g}^{k}\boldsymbol{\bf F}^{\textrm H}\circ \boldsymbol{\Phi}^{k}) + {\bf W}_{g}, \label{eq1} 
\end{aligned}
\end{equation}
where ${\bf W}_{g} = [{\bf w}_{1,g},{\bf w}_{2,g},\cdots,{\bf w}_{M,g}]^{\textrm T}$, $\boldsymbol{\Phi}^{k}\in\boldsymbol{\mathbb C}^{M \times N}$ is the phase rotation matrix between the $k$th UE and $M$ AAUs, with its $(m,n)$th element being $e^{{{j}2\pi(n-1)\varphi_{m}^{k}/{N}}}$, and `$\circ$'  denotes the Hadamard product between matrices.

As shown in (\ref{eq1}), there are $M$ unknown CFO parameters to be estimated in each $\boldsymbol{\Phi}^{k}$, and there are $M \times K$ such parameters to be estimated in the entire system, thus imposing prohibitively high computational complexity. However, we can rewrite $\varphi_{m}^{k}$ as $(\varphi_{m}^{k}-\alpha)+\alpha$, where $\alpha$ is an arbitrary value. By further inspection, $\alpha$ can be set as the CFO between the $m$th AAU and another reference AAU. Thus, for the $k$th UE, the $(m,n)$th element of $\boldsymbol{\Phi}^{k}$ comprises two information parts: the first represents the frequency difference between the $m$th AAU and a specified reference AAU, while the second contains the frequency difference between the $k$th UE and the reference AAU. Since the first part is common to all UEs, the complexity of estimating the unknown CFO parameters in $\boldsymbol{\Phi}^{k}$ can be reduced. Based on the above insights, we redefine $\boldsymbol{\Phi}^{k}$ as 
\begin{equation}
\boldsymbol{\Phi}^{k} = \boldsymbol{\Theta}^{k} {\bf D}(\varphi^{k}_{\rm sec}), \label{eq2}
\end{equation} 
where $\boldsymbol{\Theta}^{k}$ denotes the first information part, i.e., the normalized CFO among AAUs, and $\varphi^{k}_{\rm sec}$ denotes the second information part, i.e., the normalized CFO between the $k$th UE and its nearest slave AAU (i.e, its secondary AAU, as shown in Fig. \ref{system architecture}). Moreover, the $(m,n)$th element of $\boldsymbol{\Theta}^{k}$ is defined as $e^{{j2\pi(n-1)({\varphi}_{m}^{k}-\varphi_{\rm sec}^{k})}/{N}}$, where we have 
\begin{equation}
	\varphi_m^{k}-\varphi_{\rm sec}^{k}\!  =\!\varphi_m^{\rm bias}-\varphi_k^{\rm bias}+\varphi_{\textrm D}^{k,m} -(\varphi_{\rm sec}^{\rm bias}-\varphi_k^{\rm bias}+\varphi_{\textrm D}^{k,\rm sec}).  \label{eq4}
\end{equation}
In \eqref{eq4}, $\varphi_m^{\rm bias}$ and $\varphi_k^{\rm bias}$ denote the normalized CFO difference between the actual frequency and the nominal frequency of the $m$th AAU, 
and of the $k$th UE, respectively. Additionally, $\varphi_{\textrm D}^{k,m}$ and $\varphi_{\textrm D}^{k,\rm sec}$ represent the $k$th UE's normalized Doppler frequency offset relative to the $m$th AAU, and the normalized Doppler offset between the $k$th UE and its secondary AAU, respectively. Thus, the right-hand side of (\ref{eq4}) can be approximated as $\varphi_m^{\rm bias}-\varphi_{\rm sec}^{\rm bias}$ when the UE's speed is low, which means $\boldsymbol{\Theta}^{k}$ is no longer related to the UE's index, hence it can be denoted as $\boldsymbol{\Theta}$. Then, by substituting (\ref{eq2}) into (\ref{eq1}), we obtain 
\begin{equation}
{\bf Y}_g\!\approx\!\sqrt{N}\left(\sum_{k=1}^{K} \eta_{g}(\varphi^{k}_{m}){\bf H}^{k}{\bf F}_{L}^{\rm T}{\bf X}_{g}^{k}{\bf F}^{\rm H}{\bf D}(\varphi_{\rm sec}^{k})\!\right)\!\circ \boldsymbol{\Theta}\!+\!{\bf W}_{g}. \label{eq5}
\end{equation}

As seen from \eqref{eq2}, if $\boldsymbol{\Theta}$ and $\varphi_{\rm sec}^k$ were known, ${\bf \Phi}^k$ could be obtained easily. However, $\boldsymbol{\Theta}$ is distributed over different AAUs, hence it is unavailable to any individual AAU. To overcome this predicament,  we design a two-stage hierarchical frequency synchronization (HFS) scheme to collect $\boldsymbol{\Theta}$ and $\varphi_{\rm sec}^k$. More specifically, $\varphi_{\rm sec}^k$ can be obtained by the uplink synchronization in the first stage with the aid of the received signal ${\bf y}_{m,g}$ given by (\ref{y_mg}); $\boldsymbol{\Theta}$ can be obtained through the over-the-air synchronization between the master AAU (as shown in Fig. \ref{system architecture}) and the slave AAUs in the second stage, where the received signal utilized is denoted as ${\bf y}_m^{\rm syn}$ and has an expression similar to (\ref{y_mg}), obtained by letting $K=1$ and replacing the UE index $k$ with the index of an AAU in (\ref{y_mg}). 

\begin{figure}[tbp]
	\centering
	\includegraphics[width=7.6cm,height=2.9cm]{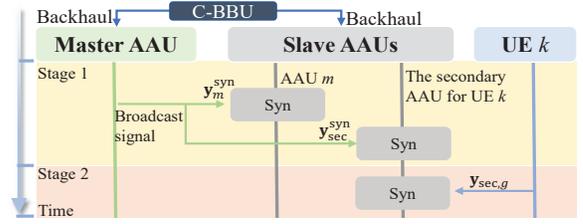}
	\caption{The process of the proposed synchronization scheme.}\label{Processdescription}
\end{figure}

The process of the proposed synchronization scheme is shown in Fig. \ref{Processdescription}, where the `Syn' block  represents any \textcolor{black}{point-to-point synchronization algorithm which is
	viable in specific application scenarios, such as the multiple signal classification (MUSIC) algorithm based scheme of~\cite{liu1998high} for OFDMA systems and the algorithms of~\cite{zhang2016blind} for SDMA systems.} Specifically, before the synchronization process starts, the master AAU is selected while the other AAUs are divided into $K$ groups of slave AAUs. Each group has a secondary AAU, which is the nearest AAU in the group for the corresponding UE. The joint action of the master AAU and secondary AAUs is like a CFO \textit{transition station}, which reduces the number of CFO parameters to be estimated from $(M\times K)$ to $(M + K)$. Additionally, to achieve better CFO estimation performance, higher SNR is preferred in both stages. Therefore, in each stage we select the specific AAU that has the lowest possible path-loss. This criterion is equivalent to choosing the AAU that is closest to the geographic centre of the area as the master AAU and the AAU that is closest to the UE considered as the secondary AAU. In particular, when the cell is small, the master AAU and the secondary AAU can be identical, which results in an even better frequency synchronization performance, as explained in the next section. 

Upon completing the AAU selection, the master AAU sends the appropriate  signals\footnote{The signals transmitted are different for blind synchronization and pilot-aided synchronization algorithms.} to the other AAUs, and the slave AAUs estimate CFOs relative to the master AAU by using any feasible synchronization algorithm. Then all the slave AAUs perform CFO compensation in the first stage. During the second stage, each secondary AAU synchronizes with its corresponding UE by using any feasible point-to-point frequency synchronization algorithm. 
As a result, $\boldsymbol{\Theta}$ and $\varphi_{\rm sec}^k$ are acquired in the first and the second stage, respectively. 

\section{Performance Analysis}
\vspace{-0.5cm}
\textcolor{black}{\subsection{System-level MSE performance}}

According to (\ref{eq2}), the estimated CFO actually includes two terms, namely $\boldsymbol{\Theta}^{k}$ and ${\bf D}(\varphi^{k}_{\rm sec})$. The MSE of $\boldsymbol{\Theta}^{k}$ depends on the first stage while that of ${\bf D}(\varphi^{k}_{\rm sec})$ depends on the second stage. Denote the MSE of the first stage and the second stage as $\sigma_1^2$ and $\sigma_2^2$, respectively. Upon considering the $(m,n)$th element of $\boldsymbol{\Phi}^{k}$ as an example, the \textit{estimated} phase can be expressed as 
\begin{equation}\label{equ12}
\begin{aligned}
\hat{\varphi}_k^m = \mathrm{Arg} [\boldsymbol{\Phi}^{k}(m,n)] = 
{\frac{\rm 2}{N}j\pi(n-1)(\varphi^{\rm stage1} + \varphi^{\rm stage2})},
\end{aligned}
\end{equation}
where ${\rm Arg}(\cdot)$ denotes the argument of a complex number, $\varphi^{\rm stage1} = \hat{ \varphi}^{\rm mas}_{\rm sec}$ is an estimate of $\varphi^k_{\rm mas} - \varphi^k_{\rm sec}$ in stage one, namely the estimated CFO between the master AAU and the secondary AAU of the $k$th UE, and $\varphi^{\rm stage2} = {\hat\varphi}^k_{\rm sec}$ is the CFO, estimated in stage two, between the $k$th UE and its secondary AAU.   
Obviously, $\varphi^{\rm stage1}-{ \varphi}^{\rm mas}_{\rm sec}$, $\varphi^{\rm stage2}-{\varphi}^k_{\rm sec}$, $\varphi_{\rm D}^{k,m}$ and $\varphi_{\rm D}^{k, \rm sec}$ are independent, and the MSE of estimating $\varphi_k^m$ by our  scheme is 
\begin{equation}
\begin{aligned}
\textcolor{black}{\mathsf {\chi}} = &{\rm E} [(\varphi^{\rm stage1}\!-\!{ \varphi}^{\rm mas}_{\rm sec}\!+\! \varphi^{\rm stage2}\!-\!{\varphi}^k_{\rm sec}\! -\!\varphi_{\rm D}^{k,m}\!+\!\varphi_{\rm D}^{k,\rm sec})^2]\\
= &2\sigma_1^2 + \sigma_2^2 + \mathrm{E}[(\varphi_{\rm D}^{k,m})^{2}]+\mathrm{E}[(\varphi_{\rm D}^{k,\rm sec})^{2}]. \label{equ13}
\end{aligned}
\end{equation}

The MSE defined in (\ref{equ13}) is a system-level metric, since it is evaluated by averaging over all the individual links between the UEs and the AAUs. Note that the frequency synchronization algorithms invoked by the individual links can be different, and their performances are related to multiple system-level factors, including the geographic distribution of AAUs and UEs, as well as the velocities and moving directions of UEs.  Therefore, to analyse the MSE performance of the proposed scheme, we have to consider the impact of the above-mentioned system-level factors. 

\begin{figure}[tbp]
	\centering
	\includegraphics[width=5cm,height=4.7cm]{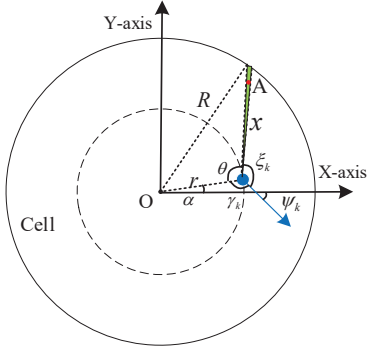}
	\caption{\textcolor{black}{Schematic diagram of MSE derivation.}} \label{proof}
\end{figure}

\textcolor{black}{To analyse the MSE of the proposed synchronization scheme, we first construct the model shown in Fig.~\ref{proof}, where we assume that the master AAU is located at the origin `O', i.e., at the center of circles. Furthermore, the secondary AAU associated with the $k$th UE is located in the green sector and represented by the red point $A$, while the blue point on the small circle represents the $k$th UE. In Fig.~\ref{proof}, we denote the distance between the $k$th UE and its master AAU as $r$, the cell radius as $R$, the moving direction of the $k$th UE as $\psi_k$, the angle between the moving direction of the $k$th UE and the connection direction of the $k$th UE to its secondary AAU as $\xi_k$. Still referring to Fig.~\ref{proof}, the angle between the connection direction of the $k$th UE to its secondary AAU and the connection direction of the $k$th UE to its master AAU is denoted as $\theta$, while the angle between the moving direction of the $k$th UE and the connection direction of the $k$th UE to its master AAU is represented by $\gamma_k$}. Additionally, we denote $\mathrm{E}[(\varphi_{\rm D}^{k,m})^{2}]$ as $\varDelta_{\rm D}^{k,m}$, $\mathrm{E}[(\varphi_{\rm D}^{k,\rm sec})^{2}]$ as $\varDelta_{\rm D}^{k,\rm sec}$, the velocity of the $k$th UE as $\upsilon_k$, and the speed of light as $c$.

\textcolor{black}{As for $\varDelta_{\rm D}^{k,m}$, it may be interpreted as the expectation of the normalized Doppler frequency offset between the $k$th UE and the master AAU, while assuming random location and velocity for the UEs. For ease of exposition, we can acquire $\varDelta_{\rm D}^{k,m}=\left(\frac{f\upsilon_k}{c\Delta f}\sin\gamma_k\right)^2$ by firstly assuming that both the location and velocity of UEs are deterministic. Furthermore, considering the randomness of the parameters, we can express the expectation of $\sin^2(\gamma_k)$ as $\int_{0}^{2\pi}\frac{1}{2\pi}\int_{0}^{2\pi} \frac{1}{2\pi}\sin^2(\gamma_k)d \theta\ d \psi_k$. As a result, when $\upsilon_k$ is determined, $\varDelta_{\rm D}^{k,m}$ can be obtained as}
\textcolor{black}{
	\begin{small}
		\begin{equation}\label{Dkm1}
		\varDelta_{\rm D}^{k,m} = \left(\frac{f\upsilon_k}{c\Delta f}\right)^2\int_{0}^{2\pi}\frac{1}{2\pi}\int_{0}^{2\pi} \frac{1}{2\pi}\sin^2(\gamma_k)d \theta\ d \psi_k, 
		\end{equation}
	\end{small}where $\Delta f$ and $f$ are the subcarrier spacing and the carrier frequency, respectively. Moreover, according to Fig. \ref{proof}, the equation $\gamma_k=2\pi-(\theta+\xi_k)$ holds. Since $\theta$ and $\xi_k$ are uniformly distributed in $(0,2\pi)$, we further obtain $\varDelta_{\rm D}^{k,m}=\frac{1}{2}\left(\frac{f\upsilon_k}{c\Delta f}\right)^2$ by substituting the equation into (\ref{Dkm1}). }

\textcolor{black}{Similarly,  $\varDelta_{\rm D}^{k,{\rm sec}}$ may be viewed as the expectation of the normalized Doppler frequency offset between the $k$th UE and its secondary AAU, while assuming random velocity and random locations for the UE and the secondary AAU. Herein we assume that the secondary AAUs are uniformly distributed in the given green sector. This assumption guarantees that the probability of the secondary AAU being in the green sector  depends on the area of the sector. Similar to the derivation of $\varDelta_{\rm D}^{k,m}$, considering the randomness of the velocity of the UE as well as the locations of the UE and the secondary AAU, 
	we formulate $\varDelta_{\rm D}^{k,{\rm sec}}$ as
	\begin{small}
		\begin{equation}\label{Dksec1}
		\varDelta_{\rm D}^{k,{\rm sec}}\!=\!\left(\frac{f\upsilon_k}{c\Delta f}\right)^2\!\int_{0}^{2\pi}\!\frac{1}{2\pi}\!\int_{0}^{R}\!\frac{1}{R}\!\int_{0}^{2\pi}\!\!\cos^2(\xi_k)p(\theta,r)d\theta\ dr\ d\psi_k,
		\end{equation}
	\end{small}where $p(\theta,r)$ is defined as the probability that the location of the secondary AAU is in the green sector of Fig. \ref{proof}. 
	Specifically, 
	for certain $r$, $p(\theta,r)d \theta$ can be expressed as the ratio of the area of the green sector to the area of the circle with radius $R$, and the expression is
	\vspace{-0.2cm}
	\begin{equation}
	p(\theta,r)d \theta=\frac{x^2}{2\pi R^2},
	\end{equation}
	where $x$ satisfies 
	\vspace{-0.2cm}
	\begin{equation}\label{x1}
	2rx\cos \theta=r^2+x^2-R^2. 
	\end{equation}
	Based on (\ref{x1}), we obtain $x=r\cos\theta+\sqrt{(R^2-r^2\sin^2\theta)}$ after a straightforward derivation. Then, by substituting $x$ and $p(\theta,r)d \theta$ into (\ref{Dksec1}) as well as adding (\ref{Dkm1}) to (\ref{Dksec1}), we formulate $\varDelta_{\rm D}^{k,m} + \varDelta_{\rm D}^{k,\rm sec}$ as (\ref{eq14}).
}

\begin{figure*}[ht] 
	\centering
\begin{small}
	\begin{equation}
	\varDelta_{\rm D}^{k,m} + \varDelta_{\rm D}^{k,{\rm sec}} = \left(\frac{f\upsilon_k}{c\Delta f}\right)^2\left\{ \frac{1}{2}+\int_{0}^{2\pi} \frac{1}{2\pi } \int_{0}^{R}\frac{1}{R} \left\{ \int_{0}^{2\pi} \frac{1}{2\pi}\left[\cos^2(\xi_k) \left(\frac{r}{R}\cos\theta+\sqrt{(1-\frac{r^2}{R^2}\sin^2\theta)}\right)\ ^2 \right] d \theta \right\} \ dr \ d \psi_k \right\} \label{eq14}
	\end{equation}
\end{small}	
\end{figure*}
However, it is challenging to convert (\ref{eq14}) into closed form. To simplify (\ref{eq14}), consider a cell-free system~\cite{Ngo2017cell}, which constitutes  an example of distributed massive MIMO scenarios and designed with the aid of the \textit{UE-centric} criterion. In such a system, wherever a UE is, there always exist AAUs nearby to serve it, \textcolor{black}{which means that the UE can be deemed to be located at the center of the circle. Thus, $\varDelta_{\rm D}^{k,{\rm sec}}$ and $\varDelta_{\rm D}^{k,m}$ degenerate into $\left(\frac{f\upsilon_k}{c\Delta f}\right)^2\int_{0}^{2\pi}\frac{1}{2\pi}\int_{0}^{2\pi} \frac{1}{2\pi}\cos^2(\psi_k-\omega)d \omega\ d \psi_k$, where $\omega$ is the angle between the connection direction of the UE to its secondary AAU and the direction of the X-axis. Then, after simple calculations} we obtain
\vspace{-0.2cm}
\begin{equation} \label{equ15}
\textcolor{black}{\mathsf {\chi}} = 2\sigma_1^2 + \sigma_2^2 + \left(\frac{f\upsilon_k}{c\Delta f}\right)^2.
\end{equation}
It is worth noting that when the master AAU and the secondary AAU are identical due to a small cell size, the MSE is improved as $\sigma_1^2 + \sigma_2^2 + \left(\frac{f\upsilon_k}{c\Delta f}\right)^2$.

\vspace{-0.5cm}
\textcolor{black}{\subsection{Computational complexity}
Let us compare the computational complexity of the proposed HFS synchronization scheme to that of the MUSIC-like algorithm based scheme in~[10] and the pairing-based efficient estimator (PBEE) of~[6]. Firstly, we analyse the overall computational complexity of the proposed scheme and of the traditional point-to-point synchronization algorithm based scheme. Upon assuming that the complexity of the specific point-to-point frequency synchronization algorithm invoked is $\mathcal{O}(\kappa)$, the overall complexity of the proposed scheme and of the traditional point-to-point synchronization algorithm based scheme is $\mathcal{O}((M+K-1)\kappa)$ and $\mathcal{O}(MK\kappa)$, respectively. In particular, when the point-to-point synchronization algorithm invoked is, for example, the MUSIC-like algorithm, we have $\kappa = L_b\rho(M+N\log_{2}N+Q) $. Then the complexity of the proposed scheme and of the MUSIC-like algorithm based scheme is $\mathcal{O}(L_{b}\rho((M-1)^2+(M+K-1)(Q+N\log_{2}N)+K))$ and $\mathcal{O}(MKL_b\rho(M+N\log_{2}N+Q))$, respectively, where $Q$ is the number of null subcarriers.  Secondly, the complexity of PBEE is $\mathcal{O} (6(\epsilon+1)KL_{b}N\lceil \frac{M}{2}\rceil (\log_{2}N+L_{b}))$ [6], where $\lceil\cdot \rceil$ is the ceiling operation. Note that $\epsilon$ and $\rho$ represent the number of searches in the one-dimensional CFO estimation of [6] and [10], which are typically selected as 50 and 140, respectively. Upon using realistic values of the system parameters involved in the above expressions, it can be demonstrated that the proposed scheme has a significantly lower complexity than both the MUSIC-like algorithm based scheme and the PBEE scheme, and the complexity reductions are mainly attributed to the hierarchical architecture.}

\begin{figure}[tbp]
	\centering
	\includegraphics[width=6.8cm,height=5.5cm]{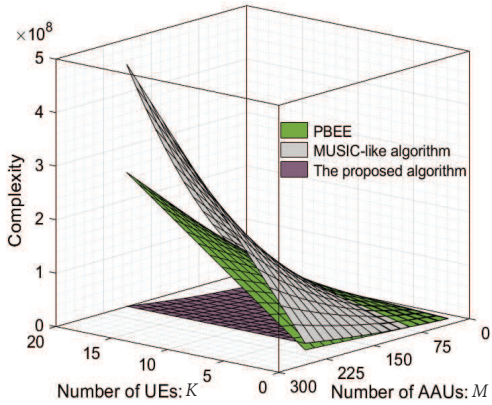}
	\caption{Computational complexity of PBEE, MUSIC-like and the proposed scheme, while assuming $N=32, P_k=20$, where $k=1,2,\cdots, K$.} \label{complexity}
\end{figure}

\vspace{-0.2cm}
\section{Simulations and Discussions}
In this section, we evaluate the proposed scheme by Monte Carlo simulations. We assume that quadrature phase shift keying (QPSK) is used,  the normalized CFO is generated with a uniform distribution in the range of $(-0.5, 0.5)$ \textcolor{black}{to avoid CFO ambiguity caused by integral part of the normalized CFO to be estimated \cite{Moose328961}}, the number of \textcolor{black}{channel taps} is $L=8$, the bandwidth of the system is $20$ MHz, the subcarrier spacing is $\Delta f =15$ kHz, the number of UEs is $K=1$,\footnote{Since OFDMA is adopted, for simplicity, in the simulations we limit our attention to the scenario of frequency synchronization for a single UE, without considering the inter-carrier interference caused by the CFO between any two UEs.} and the  transmitting power of a UE and of an AAU is $100$ mW and $500$ mW, respectively \cite{3gpp.36.101}. Since the proposed scheme is a system-level frequency synchronization approach, we assess its performance in a cell where $M$ AAUs and $K$ UEs are uniformly distributed.  
In addition, the pathloss coefficient $\beta^{m}_{k}$ between the $m$th AAU and the $k$th UE is expressed as
\vspace{-0.1cm}
\begin{equation}
\beta^{m}_{k} = \zeta - 10\lambda \log_{10}(d^{m}_{k}) + \chi^m_k, 
\end{equation}
where the pathloss exponent $\lambda$ is 3.76, the average channel gain $\zeta$ at a reference distance of $1$ km is $-148.1$ dB and the variance of shadow fading $\chi^m_k$ is 8 \cite{bjornson2017massive}. We run 50 independent trials to randomly generate the locations of UEs and AAUs by a uniform distribution, thus the impact of their locations averages out. 


In Fig. \ref{complexity}, we compare the computational complexity profiles of the three schemes considered. For the sake of fairness, we adopt the MUSIC-like algorithm in both stages of the proposed scheme. We can see that the proposed scheme has a dramatically lower computational complexity than the PBEE scheme of \cite{feng2018frequency} and the MUSIC-like algorithm based scheme of \cite{liu1998high}. This advantage is mainly attributed to the hierarchical frequency synchronization architecture of our scheme. 

\begin{figure}[tbp]
	\centering
	\includegraphics[width=7.2cm,height=5.1cm]{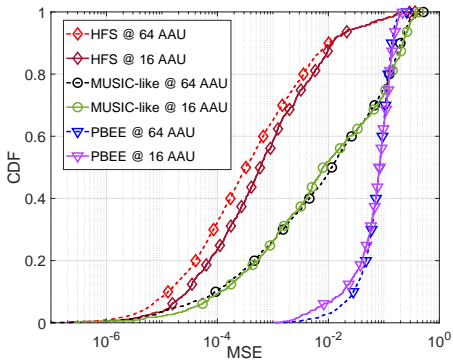}
	\caption{CDF of the MSE of the CFO estimation schemes considered, while assuming  $K=1, N=32, P_k=20, L_{b}=2$.}\label{MSEnew16and64AP}
\end{figure}

To characterize the performance of the proposed scheme, the cumulative distribution functions (CDFs) of the MSE of the proposed scheme and of the benchmark CFO estimation schemes are compared in Fig. \ref{MSEnew16and64AP} assuming both 16 AAUs and 64 AAUs. For simplicity,   
let us assume the speed of UE is $0$ m/s. 
We see that the MSE performance of the proposed scheme is orders of magnitude better than that of the MUSIC-like algorithm based scheme and that of the PBEE scheme. This is mainly because both benchmark schemes rely only on the UEs to transmit synchronization signals, which results in a low SNR at the distant AAUs.  
By contrast, the proposed scheme relies on the higher-power master AAU to broadcast the synchronization signals in the first stage, thus  attaining significant SNR improvements. Moreover, the performance advantage of the ``64 AAU" configuration over the ``16 AAU" configuration, although not large, is still more prominent in the proposed scheme than in the benchmark schemes.

\begin{figure}[tbp]
	\centering
	\includegraphics[width=8cm,height=5cm]{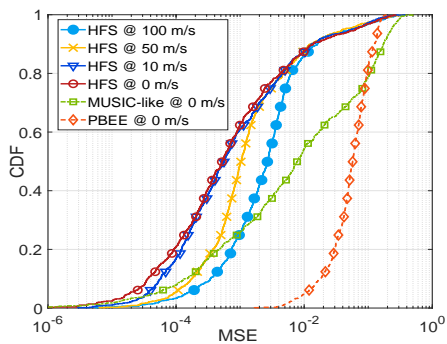}
	\caption{CDF of the MSE of the CFO estimation schemes considered under different UE speed values, while assuming $M=64, K=1, N=32, P_k=20, L_{b}=2$.}\label{VelocityComparison}
\end{figure}

In Fig. \ref{VelocityComparison}, we evaluate the impact of the Doppler frequency offset by comparing the MSE performance of the CFO estimation schemes considered under different values of UE velocity, including 0 m/s, 10 m/s, 50 m/s and 100 m/s. As expected, the higher the speed, the worse the MSE\footnote{Due to limit of space, we only show the MSE performance of the benchmark schemes under the UE speed of 0 m/s, which does not affect our conclusion drawn from the performance comparison.}, which in our scheme is due to the terms related to the Doppler frequency shift in (\ref{equ13}) and the terms related to the UE speed in (\ref{eq14}). However, the MSE of the proposed scheme assuming the UE speed of 100 m/s remains better than that of the benchmark schemes under the UE speed of 0 m/s.  

\begin{figure}[tbp]
	\centering
	\includegraphics[width=5.5cm,height=4.6cm]{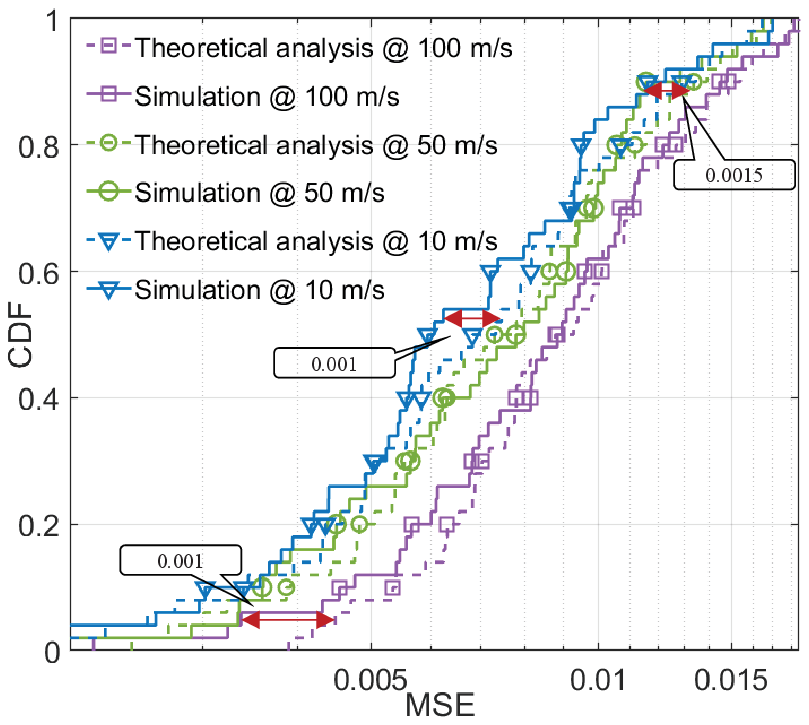}
	\caption{CDF of theoretical MSE result and numerical MSE result under different UE speed values, while assuming $M=64, K=1, N=32, P_k=20, L_{b}=2$.}\label{TheoryandSimulation}
\end{figure}

Furthermore,  in Fig. \ref{TheoryandSimulation} we compare the theoretical MSE result given by (\ref{equ15}) with the numerical MSE result for the proposed CFO estimation scheme under different UE velocity values. 
We see that the maximum theoretical prediction error is about 0.0015, as identified by the red arrow in Fig. \ref{TheoryandSimulation}, while  in most cases the prediction error is less than 0.0005. Thus, our theoretical performance analysis has a high accuracy.

\vspace{-0.2cm}
\section{Conclusion}
A high-performance yet low-complexity  hierarchical frequency synchronization scheme has been proposed for distributed massive MIMO-OFDMA systems. 
The proposed scheme relies on a two-stage hierarchical architecture, which helps significantly reduce the computational complexity. Any feasible point-to-point frequency synchronization algorithm can be embedded in the proposed scheme. 
The MSE performance of the proposed scheme 
is substantially better than that of the benchmark schemes, even if the former is used with high-mobility UEs and the latter are used with static UEs. Finally, it is demonstrated that our theoretical analysis of the MSE performance is accurate, as verified by the numerical results.


%



\ifCLASSOPTIONcaptionsoff
  \newpage
\fi

\vspace{-0.5cm}
\bibliographystyle{IEEEtran}
\bibliography{IEEEabrv,reference.bib}
\end{document}